\documentclass[journal]{IEEEtran}

\usepackage{amsmath}
\usepackage{amssymb}
\usepackage{graphicx}
\usepackage{cite}
\usepackage{algorithmic}
\usepackage{array}
\usepackage{url}
\usepackage{color}

\begin{document}

\title{Conditional Neural Bayes Ratio Estimation for Experimental Design Optimisation}

\author{S.~A.~K.~Leeney\textsuperscript{1,2},~%
T.~Gessey-Jones\textsuperscript{3},~%
W.~J.~Handley\textsuperscript{2,4},~%
E.~de~Lera~Acedo\textsuperscript{1,2},~%
H.~T.~J.~Bevins\textsuperscript{1,2},~%
and~J.~L.~Tutt\textsuperscript{1,2}%
\thanks{\textsuperscript{1}Cavendish Astrophysics, University of Cambridge, Cambridge CB3 0HE, UK.}%
\thanks{\textsuperscript{2}Kavli Institute for Cosmology, University of Cambridge, Cambridge CB3 0HA, UK.}%
\thanks{\textsuperscript{3}PhysicsX, London, UK.}%
\thanks{\textsuperscript{4}Institute of Astronomy, University of Cambridge, Cambridge CB3 0HA, UK.}%
\thanks{Corresponding author: S. A. K. Leeney (e-mail: sakl2@cam.ac.uk).}}

\markboth{IEEE Transactions on Neural Networks and Learning Systems,~Vol.~XX, No.~X, Month~Year}%
{Leeney \MakeLowercase{\textit{et al.}}: Conditional Neural Bayes Ratio Estimation for Experimental Design Optimisation}

\maketitle

\begin{abstract}
For frontier experiments operating at the edge of detectability, instrument design directly determines the probability of discovery. We introduce Conditional Neural Bayes Ratio Estimation (cNBRE), which extends neural Bayes ratio estimation by conditioning on design parameters, enabling a single trained network to estimate Bayes factors across a continuous design space. Applied to 21-cm radio cosmology with simulations representative of the REACH experiment, the amortised nature of cNBRE enables systematic design space exploration that would be intractable with traditional point-wise methods, while recovering established physical relationships. The analysis demonstrates a $\sim$20 percentage point variation in detection probability with antenna orientation for a single night of observation, a design decision that would be trivial to implement if determined prior to antenna construction. This framework enables efficient, globally-informed experimental design optimisation for a wide range of scientific applications.
\end{abstract}


\section{Introduction}
\label{sec:intro}

\IEEEPARstart{S}{ignal} detection problems pervade the physical sciences, from gravitational wave astronomy to particle physics to biomedical imaging. In many frontier experiments, the target signal is buried beneath noise and systematic effects several orders of magnitude brighter, placing the measurement at the very edge of detectability. In this sensitivity-limited regime, the design of the experiment itself, including the choice of hardware configuration, observation strategy, and data processing pipeline, becomes as critical as the underlying physics. Yet formal, statistically-motivated methods for optimising experimental design remain computationally intractable for continuous, multi-dimensional design spaces, forcing practitioners to rely on estimations, rules of thumb, or at best limited parameter sweeps when making critical design decisions. The economic stakes are immense: modern facilities such as the Square Kilometre Array (SKA), with a budget of approximately \texteuro2 billion \cite{ska2021construction}, and the Large Hadron Collider (LHC), costing approximately \$9 billion for the accelerator and detectors combined \cite{evans2008lhc}, routinely invest millions of dollars in upgraded hardware components to achieve marginal percentage-point improvements in sensitivity. In contrast, using advanced simulations to optimally select components and make design decisions upfront can yield comparable or greater gains for a fraction of the cost. This simulation-based approach addresses a massive, largely untapped gap in the design optimisation space.

Traditionally, the optimisation of experimental design has been approached through forecasting methodologies based on the Fisher information matrix \cite{tegmark1997karhunen}. While analytically powerful, the Fisher matrix formalism is fundamentally limited. It provides only local constraints on model parameters, assuming a Gaussian likelihood and posterior, which often fails to hold for complex, non-linear models. Its utility is restricted to a single, fixed point in the design space, offering no global perspective on optimisation. Full-scale simulation campaigns using Markov Chain Monte Carlo (MCMC) methods can, in principle, explore the parameter space more robustly. However, the computational cost of running MCMC analyses for every potential experimental design configuration is prohibitive, rendering a systematic exploration of the design space intractable \cite{wolz2012validity}. This creates a significant capability gap, leaving experimentalists without a computationally efficient tool to globally optimise instrument design in the face of complex, non-Gaussian uncertainties.

In recent years, Simulation-Based Inference (SBI), also known as Likelihood-Free Inference, has emerged as a powerful paradigm for parameter inference in scenarios where the likelihood function is intractable but simulations are accessible \cite{cranmer2020frontier, tejerocantero2020sbi, lueckmann2021benchmarking}. Techniques such as Neural Posterior Estimation (NPE) use normalising flows to directly model the posterior distribution \cite{papamakarios2021normalizing, greenberg2019automatic, papamakarios2019sequential, durkan2019neural}, while Neural Ratio Estimation (NRE) trains neural network classifiers to approximate the likelihood-to-evidence ratio \cite{hermans2020likelihood, miller2021truncated, delaunoy2022towards, durkan2020contrastive, benjamin2024learning}. The amortised nature of these methods, where a single trained network can perform inference on new observations in milliseconds, has driven their widespread adoption \cite{radev2020bayesflow}. While transformative for parameter inference, these methods are designed to operate on data from a \textit{fixed} experimental setup. The resulting neural estimators are conditioned only on the data, meaning that any change to the experimental design parameters (such as instrument calibration, observation duration, or hardware configuration) would necessitate a complete and costly retraining of the network. This inherent limitation prevents their direct application to the problem of global experimental design optimisation.

To bridge this gap, we introduce Conditional Neural Bayes Ratio Estimation (cNBRE), a framework that extends NRE to explicitly address the challenge of experimental design. By conditioning the neural network on both the observed data, $\mathbf{d}$, and a vector of experimental design parameters, $\alpha$, our method learns an amortised approximation of the Bayes factor, $K(\mathbf{d}, \alpha)$, across the entire design space. This single, amortised network, trained once, can then be evaluated in milliseconds to forecast the scientific return for any combination of design parameters. The cNBRE framework thereby provides a computationally efficient and globally-informed approach to optimising experimental strategy, directly linking instrument design to the probability of achieving a desired scientific outcome. This approach extends the forecasting methodology of Gessey-Jones et al.~\cite{gesseyjones2023forecasts} by explicitly conditioning the neural ratio estimator on the design parameters, allowing for amortised evaluation across the continuous design space \cite{jeffrey2023evidence}.

While cNBRE is applicable to any signal detection problem where forward simulations are available, we demonstrate its utility through a case study selected as a rigorous testing ground for the methodology: the REACH (Radio Experiment for the Analysis of Cosmic Hydrogen) experiment \cite{deleraacedo2022reach}. REACH seeks to detect the global 21-cm signal from early cosmic history, a faint radio signal with an expected amplitude that is buried beneath astrophysical foreground emission four to five orders of magnitude brighter \cite{furlanetto2006cosmology, pritchard201221cm, bowman2018absorption, singh2022saras}. Radio frequency interference (RFI) from terrestrial and satellite sources presents an additional challenge that must be mitigated \cite{leeney2023rfi, anstey2024rfi}, as do effects such as lunar occultation of the foreground sky \cite{pattison2025occultation}. This is arguably one of the most extreme signal-to-noise problems in contemporary science, making it an ideal testing ground for cNBRE. Crucially, the REACH collaboration has developed mature, physics-based simulation infrastructure that captures the complex instrumental systematics and environmental effects dominating this detection problem \cite{anstey2021informing, pattison2025environmental}. This combination of extreme sensitivity requirements and the availability of sophisticated forward models makes REACH an ideal testbed for demonstrating the capabilities of cNBRE in high-complexity scientific settings.

This paper details the theory and application of cNBRE for experimental design. We begin by outlining the statistical foundations of Bayesian model selection and Neural Ratio Estimation. We then develop the cNBRE formalism, demonstrating how conditioning on design parameters enables amortised forecasting. We apply this framework to the REACH 21-cm experiment, conducting two key forecasting experiments. First, as a validation of our method, we use the total observation time, $\tau$, as the design parameter. The cNBRE forecast correctly recovers the expected sigmoid relationship between observation time and detection probability, consistent with radiometer equation scaling. Second, as a proof-of-concept for design optimisation, we investigate the impact of the instrument's physical orientation angle. Our results reveal a non-trivial, $\sim$20 percentage point variation in the 21-cm signal detection probability as a function of orientation, demonstrating the framework's ability to uncover crucial design dependencies that would be missed by traditional methods. We conclude by discussing the broad applicability of this amortised forecasting approach for a wide range of problems in the physical sciences.

\section{Theory}\label{sec:theory}

The optimisation of experimental design for signal detection fundamentally relies on the ability to quantify the distinguishability between competing physical models. In this work, we adopt a Bayesian model selection framework \cite{jeffreys1961theory, kass1995bayes}, comparing a signal model, $M_1$ (e.g., global 21-cm signal plus foregrounds and noise), against a null model, $M_0$ (foregrounds and noise only). The relative evidence for these models given an observed data vector $\mathbf{d}$ is quantified by the Bayes factor $K$:
\begin{equation}
    K(\mathbf{d}) = \frac{p(\mathbf{d}|M_1)}{p(\mathbf{d}|M_0)} = \frac{\int p(\mathbf{d}|\theta_1, M_1) p(\theta_1|M_1) \mathrm{d}\theta_1}{\int p(\mathbf{d}|\theta_0, M_0) p(\theta_0|M_0) \mathrm{d}\theta_0},
\end{equation}
where $\theta_m$ represents the parameter vector for model $M_m$ (with $m \in \{0, 1\}$), and $p(\theta_m|M_m)$ is the corresponding prior. Calculating these marginal likelihood integrals is computationally expensive, particularly for high-dimensional parameter spaces, rendering standard nested sampling \cite{skilling2006nested, ashton2022nested, buchner2023nested} or MCMC \cite{trotta2008bayes} approaches intractable for optimising experimental design where this evaluation must be repeated for thousands of configurations.

\subsection{Neural Bayes Ratio Estimation}

To overcome this computational bottleneck, we utilise Neural Bayes Ratio Estimation (NBRE), a simulation-based inference technique that reformulates the calculation of the Bayes factor as a classification problem \cite{hermans2020likelihood, miller2021truncated, benjamin2024learning}. This approach builds on the foundational insight that optimal binary classifiers learn monotonic functions of the likelihood ratio \cite{cranmer2020frontier}.

We define a binary classification task where a neural network $f_\phi(\mathbf{d})$ with trainable parameters $\phi$ is trained to distinguish between data generated from $M_1$ and $M_0$. Let $m \in \{0, 1\}$ denote the model label. We generate a training dataset $\{\mathbf{d}_i, m_i\}_{i=1}^N$ by sampling equally from both models, implying equal prior odds $p(M_1) = p(M_0) = 0.5$.

The network is trained to minimise the expected exponential loss. The exponential loss function is defined as:
\begin{equation}\label{eq:exp_loss}
    \mathcal{L}[\phi] = \sum_{i=1}^N \ell(m_i, f_\phi(\mathbf{d}_i)),
\end{equation}
with the pointwise loss:
\begin{equation}
    \ell(m, r) = \begin{cases}
      \exp(-r/2) & \text{if } m = 1 \\
      \exp(r/2) & \text{if } m = 0.
   \end{cases}
\end{equation}
The exponential loss provides a direct mapping to the log-Bayes factor, avoiding potential numerical issues with sigmoid saturation that can occur with binary cross-entropy when Bayes factors span many orders of magnitude. Alternative loss functions are possible for neural ratio estimation \cite{jeffrey2023evidence}; we experimented with several and found the exponential loss to be preferable for our application.

In the limit of infinite training data, minimising this loss corresponds to minimising the functional:
\begin{equation}\label{eq:continuum_loss}
    \mathcal{J}[f] = \int \left[ p(\mathbf{d}|M_1) e^{-f(\mathbf{d})/2} + p(\mathbf{d}|M_0) e^{f(\mathbf{d})/2} \right] \mathrm{d}\mathbf{d}.
\end{equation}
Taking the functional derivative with respect to $f(\mathbf{d})$ and setting it to zero yields the optimality condition:
\begin{equation}
    \frac{\delta \mathcal{J}}{\delta f} = -\frac{1}{2} p(\mathbf{d}|M_1) e^{-f^*(\mathbf{d})/2} + \frac{1}{2} p(\mathbf{d}|M_0) e^{f^*(\mathbf{d})/2} = 0.
\end{equation}
Rearranging these terms, we find:
\begin{equation}
    e^{f^*(\mathbf{d})} = \frac{p(\mathbf{d}|M_1)}{p(\mathbf{d}|M_0)} = K(\mathbf{d}).
\end{equation}
Thus, the optimal network output is the log-Bayes factor, $f^*(\mathbf{d}) = \log K(\mathbf{d})$. Once trained, the network provides a fast, amortised estimator of the Bayes factor for any input data $\mathbf{d}$.

\subsection{Conditional Extension}

To enable experimental design optimisation, we extend the NBRE framework to include design parameters. Let $\alpha$ denote the vector of controllable experimental design parameters (e.g., integration time, antenna orientation), distinct from the inferred physical parameters $\theta$. The generative process now depends on $\alpha$, yielding the conditional likelihood $p(\mathbf{d}|\theta, \alpha, M)$.

We introduce Conditional Neural Bayes Ratio Estimation (cNBRE) by conditioning the network on the design parameters, $f_\phi(\mathbf{d}, \alpha)$. The training set is augmented to triplets $\{\mathbf{d}_i, m_i, \alpha_i\}$, where $\alpha_i$ is sampled from a proposal distribution $p(\alpha)$ covering the design space of interest. Crucially, by sampling nuisance parameters $\theta$ (foregrounds, cosmological parameters) from their priors $p(\theta)$ during training, the network implicitly performs the marginalisation integral. The resulting output approximates the ratio of \textit{marginal} likelihoods rather than a point-wise likelihood ratio. The network learns to approximate the conditional log-Bayes factor:
\begin{equation}
    f_\phi(\mathbf{d}, \alpha) \approx \log K(\mathbf{d}|\alpha) = \log \frac{p(\mathbf{d}|\alpha, M_1)}{p(\mathbf{d}|\alpha, M_0)}.
\end{equation}
This approach yields a single amortised network capable of predicting the Bayes factor for any data realisation and any experimental configuration within the training range \cite{cranmer2020frontier, gesseyjones2023forecasts, jeffrey2023evidence}. This eliminates the need to retrain the network for each new design choice, accelerating the optimisation process by orders of magnitude.

\subsection{Detection Probability}

The primary objective of experimental design in this context is to maximise the probability of confidently detecting the signal if it exists. Unlike standard Bayesian Experimental Design approaches \cite{lindley1956measure, chaloner1995bayesian} (including recent variational methods that enable gradient-based optimisation of Expected Information Gain \cite{foster2019variational}, neural mutual information estimation \cite{kleinegesse2020bayesian}, and implicit adaptive design \cite{ivanova2021implicit}), we optimise a decision-theoretic utility: the probability of decisive model selection. In the ``discovery'' regime of signal detection, the priority is establishing the \textit{presence} of a signal over constraining its parameters. This distinction is crucial: cNBRE is designed for hypothesis testing (Is there a signal?) rather than parameter estimation (What are the signal's properties?). We define the detection probability $P_{\text{det}}(\alpha)$ as the probability that the Bayes factor exceeds a specified significance threshold $K_{\text{crit}}$ given that $M_1$ is true:
\begin{equation}
    P_{\text{det}}(\alpha) = \int \mathbb{I}[K(\mathbf{d}|\alpha) > K_{\text{crit}}] \, p(\mathbf{d}|\alpha, M_1) \, \mathrm{d}\mathbf{d},
\end{equation}
where $\mathbb{I}[\cdot]$ is the indicator function.

In practice, we estimate $P_{\text{det}}(\alpha)$ via Monte Carlo sampling. We simulate a batch of $J$ data realisations $\{\mathbf{d}_j\}_{j=1}^J$ from the signal model $M_1$ with design $\alpha$, and compute the empirical detection rate using the trained cNBRE network:
\begin{equation}
    \hat{P}_{\text{det}}(\alpha) \approx \frac{1}{J} \sum_{j=1}^J \mathbb{I}[f_\phi(\mathbf{d}_j, \alpha) > \log K_{\text{crit}}].
\end{equation}
Since the indicator function is non-differentiable, we cannot maximise $\hat{P}_{\text{det}}(\alpha)$ using standard gradient descent through the indicator. Instead, optimisation over $\alpha$ is performed by evaluating the amortised estimator across a grid of design parameters or using derivative-free optimisation methods.

The threshold $K_{\text{crit}}$ corresponds to a desired statistical significance level (in units of standard deviations, $\sigma$). This mapping assumes an approximately Gaussian distributed test statistic. Although Bayes factors are the more principled Bayesian measure of evidence, we present results in terms of $\sigma$ levels as they are more widely understood by physical scientists. The approximate correspondence between $\sigma$ levels and Bayes factors is given in Table~\ref{tab:sigma_levels} \cite{kass1995bayes}. We note that even modest changes in design parameters can shift the detection probability across these thresholds; for example, the $\sim$20 percentage point variation found in Section~\ref{sec:results} spans the difference between a marginal and a confident detection.

\begin{table}[h]
    \centering
    \caption{Approximate correspondence between statistical significance levels and Bayes Factors.}
    \label{tab:sigma_levels}
    \begin{tabular}{ccc}
        \hline
        Significance & $K$ & $\log K$ \\
        \hline
        $1\sigma$ & $\sim 3$ & $\sim 1.1$ \\
        $2\sigma$ & $\sim 12$ & $\sim 2.5$ \\
        $3\sigma$ & $\sim 150$ & $\sim 5.0$ \\
        $5\sigma$ & $\sim 1.7\times 10^6$ & $\sim 14.4$ \\
        \hline
    \end{tabular}
\end{table}

\section{Methods}\label{sec:methods}

\subsection{Model Definitions}
We formulate the detection problem as a Bayesian model comparison between a null model $M_0$ (foregrounds and noise only) and a signal model $M_1$ (global 21-cm signal, foregrounds, and noise). The data vector $\mathbf{d}$ is modelled as:
\begin{align}
    M_0: \quad \mathbf{d} &= \mathbf{d}_{\text{fg}}(\theta_{\text{fg}}) + \mathbf{n}, \\
    M_1: \quad \mathbf{d} &= \mathbf{d}_{\text{fg}}(\theta_{\text{fg}}) + \mathbf{d}_{\text{21}}(\theta_{\text{21}}) + \mathbf{n}.
\end{align}
The noise term $\mathbf{n}$ is modelled as uncorrelated, heteroscedastic Gaussian noise, with variance determined by the frequency-dependent system temperature $T_{\text{sys}}(\nu)$. The standard deviation in each frequency channel is given by the radiometer equation:
\begin{equation}
    \sigma_n(\nu_i) = \frac{T_{\text{sys}}(\nu_i)}{\sqrt{\Delta\nu \tau}},
\end{equation}
where $\Delta\nu$ is the channel bandwidth and $\tau$ is the integration time. Achieving the precision required to detect the faint 21-cm signal necessitates careful radiometer calibration \cite{murray2022bayesian, leeney2025radiometer, kirkham2025noise, kirkham2025drift}. Note that $\tau$ plays a dual role: it is an intrinsic parameter defining the noise level in the data generation process, and a controllable experimental design parameter we seek to optimise.

The target cosmological signal, $\mathbf{d}_{\text{21}}$, is modelled as a phenomenological Gaussian absorption profile \cite{cohen2017charting} centred at frequency $\nu_0$ with amplitude $A$ and width $w$:
\begin{equation}
    T_{21}(\nu) = -A \exp\left( -\frac{(\nu - \nu_0)^2}{2w^2} \right).
\end{equation}

\subsection{The REACH Forward Model}
To simulate realistic observations, we employ the REACH forward model \cite{deleraacedo2022reach, anstey2021informing}. We implemented a high-performance, fully differentiable version of this pipeline using JAX \cite{bradbury2018jax}, a numerical computation library that enables automatic differentiation and XLA compilation. This JAX implementation reduces the computational cost of the forward model by over two orders of magnitude compared to the original CPU-based pipeline, from hundreds of CPU-years to approximately two GPU-days for a comparable analysis \cite{tutt2026gpu}. This acceleration provides two key advantages: (1) GPU-accelerated generation of mock data fast enough to simulate data on-the-fly during training; and (2) end-to-end differentiability, preserving gradient flow from the loss function through the entire simulation pipeline. The resulting data vector $\mathbf{d}$ consists of 151 frequency channels spanning the bandwidth 50-200 MHz.

The forward model is characterised by a large number of nuisance parameters that make this detection problem particularly challenging. First, the target signal is buried beneath astrophysical foreground emission from our own Galaxy that is $10^4$--$10^5$ times brighter. This foreground emission is modelled using a physically motivated sky average temperature of the form $T_{\text{sky}} = \sum_i a_i \nu^{b_i}$, multiplied by a beam chromaticity factor to account for the variation in antenna response as a function of frequency. The beam chromaticity is modelled using beam factors as described in \cite{anstey2021informing, sims2023bayesian, hibbard2023fitting}. Second, the antenna's sensitivity varies with frequency (``chromaticity''), coupling the bright foregrounds into spectral structures that can mimic the target signal. In this work, the antenna beam chromaticity is accounted for in the foreground model and depends on the physical design parameters $\alpha$ we seek to optimise. Third, receiver noise and gain variations introduce additional spectral structure that must be modeled.
The key point for cNBRE is that these nuisance parameters create a high-dimensional space over which the Bayes factor must be marginalized: precisely the computational bottleneck that motivates our approach. Priors for the signal model parameters (amplitude, centre frequency, and width) are physically motivated distributions learned from $\sim$22,000 astrophysical simulations \cite{bevins2021globalemu} via the \textsc{margarine} framework \cite{bevins2022margarine}, covering amplitudes of $\sim$0.1--300~mK, centre frequencies of $\sim$40--140~MHz, and widths of $\sim$2--30~MHz. This ensures the training data spans a physically plausible range of scenarios.

\subsection{Network Architecture and Training}
We employ a split MLP network architecture. The network takes in 151 frequency channels and compresses them through layers of size 96, 48, 24, and 12. This representation is then mixed with the experimental design parameters $\alpha$ in a subsequent MLP section with three hidden layers of 100 neurons each. We performed light optimisation over the neuron count but found this was not a critical design parameter. We found this architecture led to faster convergence compared to the standard MLP initially considered. This late-fusion conditioning strategy allows the network to learn how the same spectral features should be weighted differently depending on the experimental configuration. This architectural choice is physically motivated: design parameters like integration time primarily affect the global signal-to-noise ratio or the relative weighting of the entire band, which are global properties most effectively modulated after local spectral features have been extracted.

A critical advantage of implementing the forward model in JAX is that the entire pipeline, from simulation through network evaluation, is fully differentiable with respect to all model and design parameters via automatic differentiation, and can be compiled for GPU acceleration. This enables on-the-fly data generation during training: rather than pre-computing a fixed dataset, we generate fresh training samples at each iteration. We maximise the batch size to fit on the GPU (single A100 40GB) and run for 5 million iterations, monitoring the loss on a held-out validation set to confirm convergence. Since each iteration draws fresh samples from the simulator, the network never sees the same data realisation twice, effectively eliminating overfitting by construction.

The network minimises the exponential loss (\ref{eq:exp_loss}) using the AdamW optimiser \cite{loshchilov2019adamw}. We followed the optimisation procedure outlined in \cite{tuningplaybook}. Adding nuisance parameters to the network as input (while excluding signal parameters) was found to speed up convergence by an order of magnitude. We tried many optimiser configurations and found fixed warmup followed by cosine decay led to optimal convergence. We ran training on many random seeds to confirm that our results are robust to random initialisation \cite{leeney2024randomness}. We apply global dataset-wise normalisation: a single mean and standard deviation are computed over a large subset of training data and applied to input vectors. Training took approximately 4 hours for the noise experiment and 10 hours for the antenna experiment.

\subsection{Experimental Design Parameters}
We conduct two forecasting experiments to demonstrate the cNBRE framework. For each experiment, a separate network was trained with the relevant design parameter $\alpha$. In Experiment 1 (Integration Time), we vary the total observation time $\tau$ across a log-uniform range from $10^{-3}$ to $10^3$ minutes, with training samples drawn from $\tau \sim \text{LogUniform}(10^{-3}, 10^3)$ min. In Experiment 2 (Antenna Orientation), we fix $\tau = 10$ hours and vary the antenna azimuth angle $\psi$ from $0^\circ$ to $360^\circ$ at a fixed Local Sidereal Time (LST = 0 h), with training samples drawn from $\psi \sim \text{Uniform}(0^\circ, 360^\circ)$.

\subsection{Evaluation Protocol}
To generate the detection probability surfaces, we use quasi-random uniform sampling to fill a grid of the parameters of interest. For each design configuration $\alpha$, we generate samples from the signal model $M_1$ and compute the detection probability $P_{\text{det}}(\alpha)$ by calculating the fraction of samples where the predicted Bayes factor exceeds the theoretical significance thresholds ($1\sigma$, $3\sigma$, $5\sigma$) defined in Table~\ref{tab:sigma_levels}.

\subsection{Validation Against Nested Sampling}
To ensure the robustness and accuracy of the cNBRE framework, we performed a direct validation against nested sampling \cite{skilling2006nested, ashton2022nested, buchner2023nested} using \cite{yallup2025nested}, a gold-standard approach for Bayesian evidence estimation. Individual data realisations were randomly drawn from the same distributions as used for the cNBRE training, sampling from both the signal model ($M_1$) and the null model ($M_0$) with nuisance parameters drawn from their respective priors. For each simulated data realisation, we executed independent nested sampling runs using, computing the Bayesian evidence for both models. The Bayes factor was then calculated as the ratio of these evidences, $K(\mathbf{d}) = p(\mathbf{d}|M_1) / p(\mathbf{d}|M_0)$. These gold-standard values were directly compared against the amortised Bayes factors predicted by the trained cNBRE network for the identical data realisations, allowing us to assess the calibration and consistency of the estimator.

\section{Results}\label{sec:results}

We apply the cNBRE framework to the REACH experiment to demonstrate its utility for experimental design optimisation. We present two forecasting experiments: a study using observation time as the design parameter, and a proof-of-concept optimisation of the antenna orientation. We then validate the framework by comparing its Bayes factor estimates against traditional nested sampling.

\subsection{Experiment 1: Observation Time}

In this first experiment, we validate that the cNBRE framework recovers the expected physical relationship between signal detectability and noise levels. We vary the integration time $\tau$ from $10^{-3}$ to $10^3$ minutes, which directly controls the radiometric noise level $\sigma_n \propto \tau^{-1/2}$.

\begin{figure}[t]
\centering
\includegraphics[width=\columnwidth]{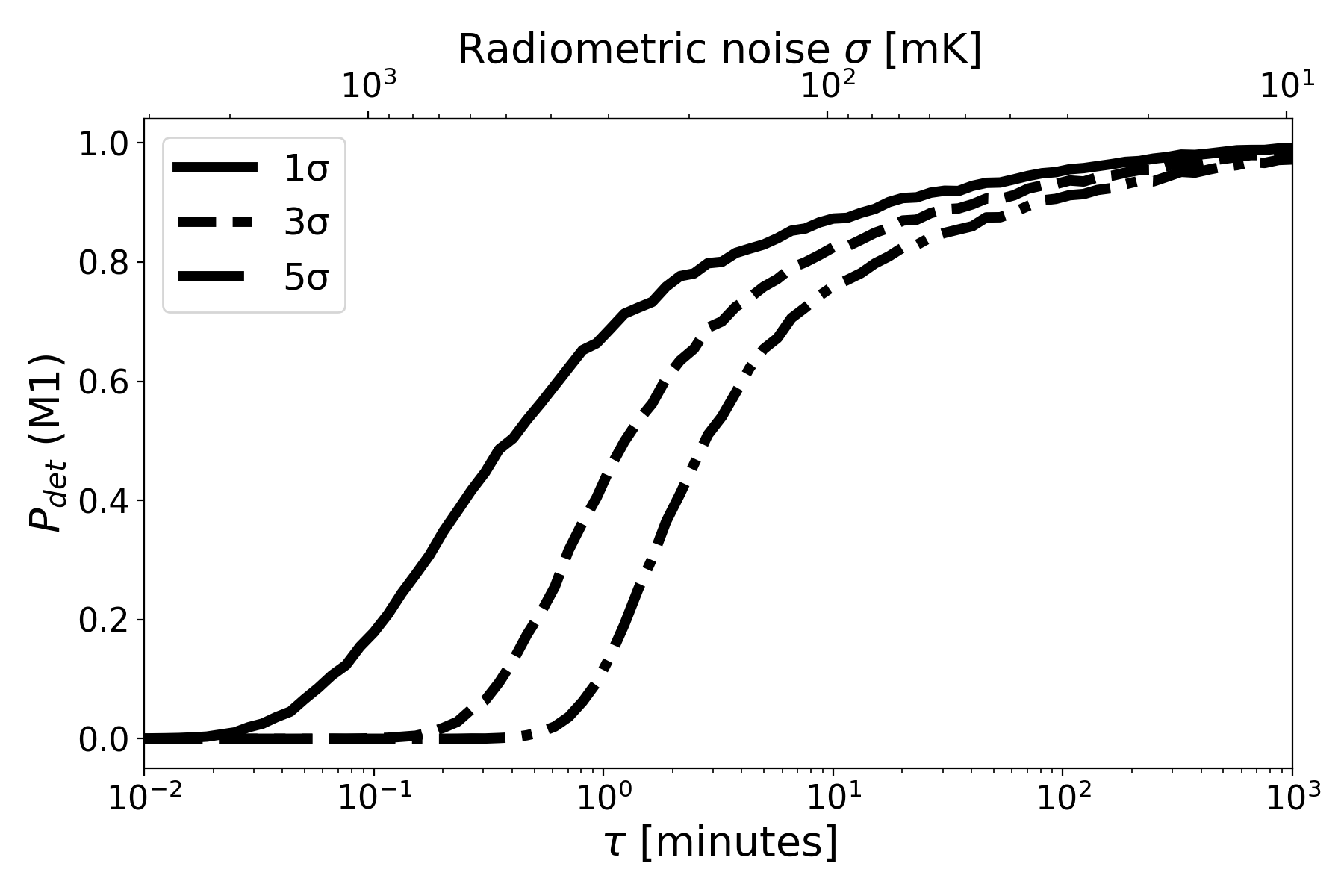}
\caption{Detection probability as a function of observation time $\tau$ for three Bayes factor thresholds corresponding to $1\sigma$ ($K\approx 3$), $3\sigma$ ($K\approx 150$), and $5\sigma$ ($K\approx 1.7\times 10^6$) significance.}
\label{fig:tau_vs_probs}
\end{figure}

Figure~\ref{fig:tau_vs_probs} displays the detection probability $P_{\text{det}}(\tau)$ for three significance thresholds: weak evidence ($K \approx 3$, corresponding to $\sim 1\sigma$), strong evidence ($K \approx 150$, $\sim 3\sigma$), and decisive evidence ($K \approx 1.7\times 10^6$, $\sim 5\sigma$).

The results exhibit a monotonic sigmoid profile, marginalised over all physical signal parameters and nuisance parameters. At short observation times ($\tau \lesssim 1$~minute) the radiometric noise dominates the data and detection probability remains near zero; at long observation times ($\tau \gtrsim 10$~minutes) the noise averages down sufficiently for the signal to emerge, and $P_{\mathrm{det}} \to 1$. Between these two limiting regimes lies a transition region whose location depends on the chosen significance threshold: the $1\sigma$ curve rises first, requiring less integration time, while the $3\sigma$ and $5\sigma$ curves are shifted to progressively longer observation times. This behaviour is consistent with expectations from the radiometer equation, where the noise level scales as $\tau^{-1/2}$, and confirms that the cNBRE estimator has learned the correct physical scaling.

Detection probabilities are estimated via Monte Carlo sampling with 10,000 realisations per design point, yielding statistical uncertainties of order 1\%.

\begin{figure*}[t]
\centering
\includegraphics[width=\textwidth]{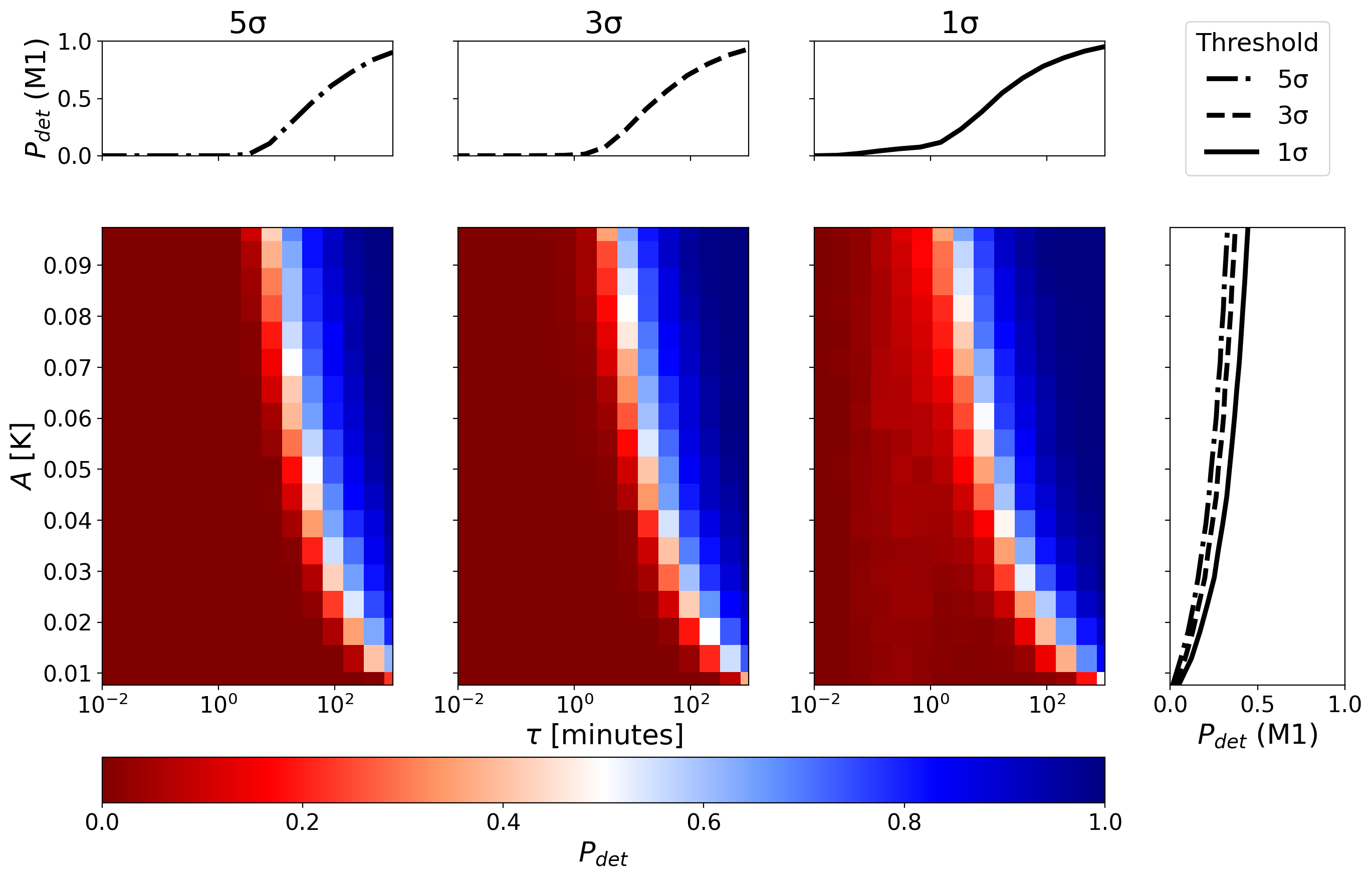}
\caption{Detection probability as a function of observation time ($\tau$) and signal amplitude ($A$). Detection probabilities are computed on test sets with signals injected at fixed amplitudes; the network marginalizes over the full signal prior during training. The panels show $5\sigma$, $3\sigma$, and $1\sigma$ thresholds, confirming that detection probability improves with longer integration and larger amplitudes.}
\label{fig:amp_tau_panels}
\end{figure*}

To further elucidate the dependence of detectability on signal parameters, Figure~\ref{fig:amp_tau_panels} presents a detailed breakdown of the detection probability across the observation time ($\tau$) and signal amplitude ($A$) plane, displaying heatmaps for $5\sigma$, $3\sigma$, and $1\sigma$ thresholds. The results confirm the expected physical trend where detection probability rises with both longer integration times and larger signal amplitudes, allowing even weak signals ($\sim$0.01~K) to be detected at lower thresholds. The marginal plots indicate a relatively uniform response across the amplitude range, demonstrating the cNBRE framework's consistent forecasting capability throughout the parameter space.

\begin{figure*}[t]
\centering
\includegraphics[width=\textwidth]{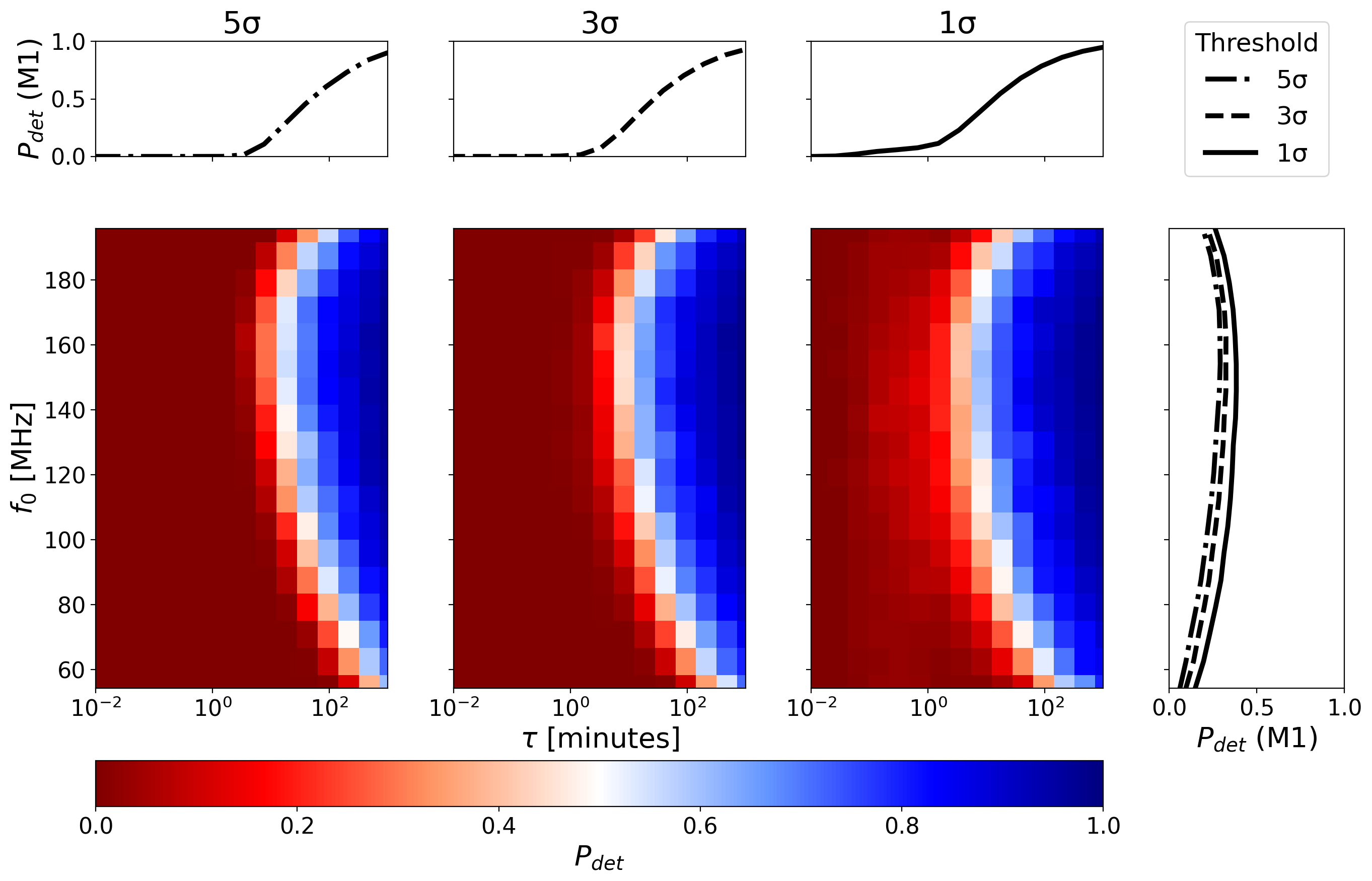}
\caption{Multi-panel visualisation of detection probability as a function of observation time ($\tau$) and signal central frequency ($f_0$). Detection probabilities are computed on test sets with signals injected at fixed $f_0$ values; the cNBRE network itself marginalizes over the full signal prior during training. The three heatmaps show $5\sigma$, $3\sigma$, and $1\sigma$ thresholds (left to right). Marginal plots highlight reduced sensitivity at the band edges.}
\label{fig:f0_tau_panels}
\end{figure*}

To further characterise the detection landscape, Figure~\ref{fig:f0_tau_panels} extends this analysis to the signal central frequency ($f_0$). The heatmaps illustrate that signals in the 100--140 MHz band are more readily detected, whereas sensitivity degrades significantly at the band edges (60 MHz and 190 MHz) regardless of integration time. Marginal plots confirm that while longer observations universally improve detection prospects, the spectral location of the signal remains a dominant factor in determining the achievable significance. It is particularly interesting to observe how physics-based parameters, such as the signal depth, directly correspond to instrument parameters like the required observation time to reach a given significance.

These two analyses (central frequency and amplitude) are representative examples of how the amortised cNBRE framework enables exploration of the detection landscape. A key advantage of amortization is that once the network is trained, such parameter slices can be generated at negligible additional computational cost; the same trained network can produce arbitrarily many such visualisations. These analyses confirm that while observation time is the primary driver of sensitivity, the specific spectral characteristics of the signal, particularly its central frequency, play a significant secondary role.

\subsection{Experiment 2: Antenna Orientation (Proof-of-Concept)}

As a proof-of-concept for non-trivial design optimisation, we investigate the impact of the antenna's physical orientation on detection probability \cite{cumner2022antenna, cumner2024antenna}. We vary the antenna azimuth angle for several observations over a single night.

\begin{figure}[t]
\centering
\includegraphics[width=\columnwidth]{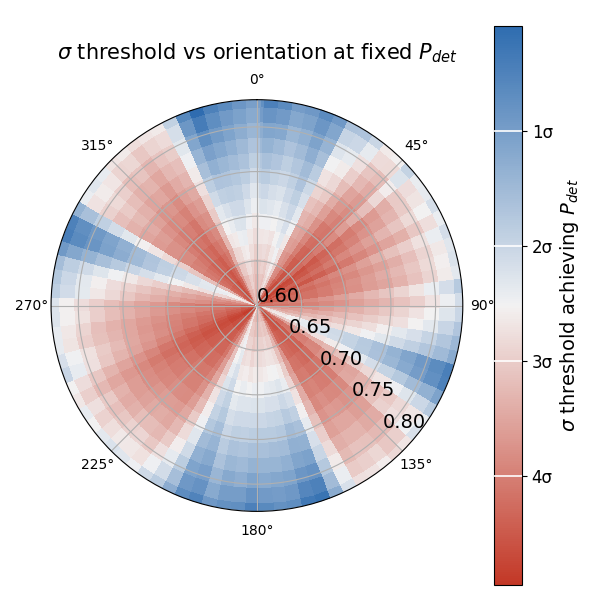}
\caption{Polar heatmap showing the variation in detection probability (radial axis) and achievable significance (colour scale) as a function of antenna orientation angle (angular axis) for several observations over a single night. The four-fold symmetry reflects the interaction of the rectangular dipole beam with galactic foregrounds.}
\label{fig:orientation_heatmap}
\end{figure}

Figure~\ref{fig:orientation_heatmap} shows the detection probability as a function of orientation angle. The polar heatmap reveals a distinct four-fold symmetry. This pattern arises from the interaction between the rectangular dipole geometry and the relatively smooth foreground structure in this sky snapshot. The orthogonal symmetry axes correspond to the principal axes of the antenna beam pattern.

This analysis indicates a significant performance differential between optimal and suboptimal orientations. We observe a $\sim$20 percentage point variation in detection probability for the $3\sigma$ threshold for a single night of observation. While this specific result does not account for the variation in sky structure over a full year of observations, it demonstrates the capability of the cNBRE framework to identify complex, non-linear dependencies in the design space that would be computationally prohibitive to explore with standard MCMC methods.

\subsection{Validation Against Nested Sampling}

\begin{figure}[t]
\centering
\includegraphics[width=\columnwidth]{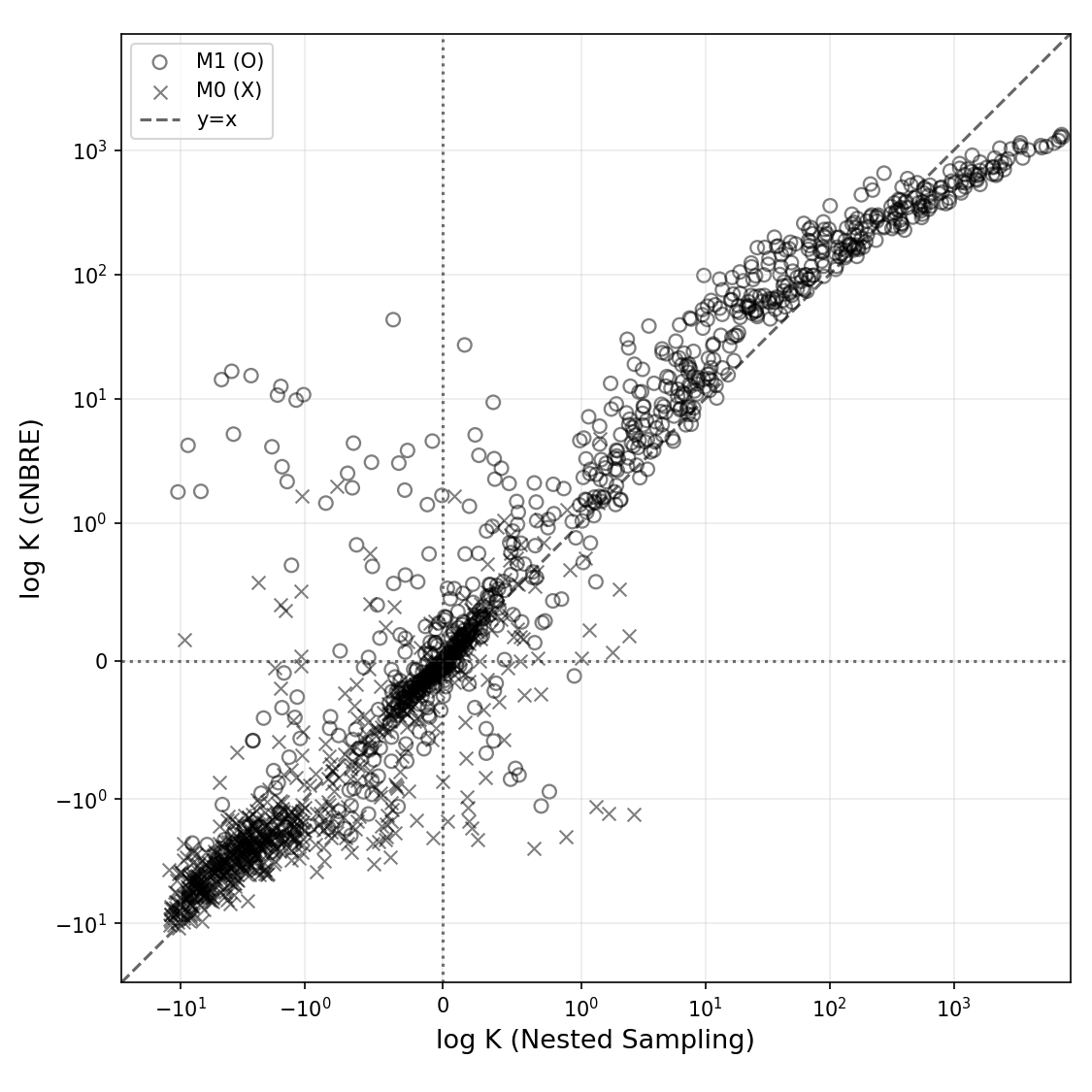}
\caption{Comparison of log-Bayes factors estimated by cNBRE (y-axis) versus nested sampling (x-axis) for 1,600 simulated data realisations (811 $M_1$, 789 $M_0$). Circles denote $M_1$ samples (signal present); crosses denote $M_0$ samples (null model). The dashed line indicates perfect agreement. Both methods show broad agreement in the decision-relevant regime, with $M_0$ samples clustering at negative $\log K$ and $M_1$ samples at positive $\log K$ for both estimators. At very high Bayes factors the cNBRE estimates saturate below the nested sampling values, a consequence of classifier output saturation that is expected and practically irrelevant for detection decisions.}
\label{fig:cnbre_ns}
\end{figure}

To validate that the cNBRE framework produces calibrated Bayes factor estimates, we compare against traditional nested sampling following the methodology described in Section~\ref{sec:methods}. Figure~\ref{fig:cnbre_ns} presents a scatter plot comparing log-Bayes factors computed via cNBRE against those obtained from full nested sampling runs on 1,600 identical simulated data realisations (811 from the signal model $M_1$ and 789 from the null model $M_0$).

The results show broad agreement between the two methods across the decision-relevant regime. For both $M_1$ (circles) and $M_0$ (crosses) samples, the estimated log-Bayes factors cluster around the identity line, confirming that the cNBRE network produces calibrated estimates consistent with the nested sampling gold standard. At very high Bayes factors ($\log K \gtrsim 10$), the cNBRE estimates begin to saturate below the nested sampling values. This tail-off is expected: the classifier's output saturates near unity for overwhelming detections, compressing the implied log-Bayes factor. Crucially, this deviation occurs in a regime where the precise value is immaterial for detection decisions.

A notable asymmetry is visible in the null-model samples: nested sampling assigns a non-negligible fraction of $M_0$ realisations positive log-Bayes factors, whereas cNBRE consistently assigns them negative values. A Bayesian McNemar test on the 1,600 paired classifications finds strong evidence that cNBRE outperforms nested sampling on discordant samples ($P(p>0.5 \mid \mathrm{data}) > 0.99$). This suggests that the amortised nature of cNBRE, which learns the decision boundary from the full distribution of data under both models rather than fitting each realisation independently, may confer a natural robustness against false positives. Exploring the utility of cNBRE as a null-testing framework in decision-critical regimes is a promising direction for future work.

\section{Discussion}\label{sec:discussion}

\subsection{Methodological Contribution}
We have presented Conditional Neural Bayes Ratio Estimation (cNBRE) as a framework for global experimental design optimisation. Unlike standard SBI approaches which are fixed to a single experimental configuration, cNBRE learns the global dependence of the Bayes factor on the instrument design. This approach complements traditional intuition-driven design processes with quantitative statistical rigor, enabling the identification of non-trivial optima that may be counter-intuitive or inaccessible to linear forecasting methods.

A critical methodological advantage of cNBRE over traditional black-box optimisation is differentiability. Because the estimator is a neural network, it provides a differentiable surrogate of the utility surface. This allows optimisation of design parameters $\alpha$ by following the gradients of the learned Bayes factor, a differentiable proxy for detection probability, effectively bypassing the non-differentiability of the forward simulator. This property, while not exploited in the present work, opens the door to gradient-based optimisation algorithms (e.g., stochastic gradient ascent) that could directly maximise scientific return in high-dimensional design spaces (such as the complex geometric shape of an antenna) where exhaustive grid searches are computationally prohibitive (see Section~\ref{sec:future} for further discussion). In the present study, we focus on demonstrating the amortised landscape characterisation capability: the ability to map the complete detection surface from a single trained model. This capability allowed us to map the full sigmoid profile of detection probability as a function of observation time (Fig.~\ref{fig:tau_vs_probs}) and the non-monotonic dependence of detection probability on antenna orientation, revealing a four-fold symmetry that a coarser linear forecast might miss.

\subsection{Computational Considerations}
An advantage of the cNBRE framework is its computational efficiency during the evaluation phase. The initial training phase requires approximately 4--10 GPU-hours on a single NVIDIA A100, a one-time amortisation cost. This training phase is many orders of magnitude cheaper than the alternative of running independent sampling-based evidence evaluations for every point in the design space. Once the network is trained, evaluating the scientific return (detection probability) for any new design configuration takes only milliseconds.

To contextualise this computational trade-off, we compare against the baseline of performing a standard nested sampling run for each individual design configuration $\alpha$. In our parameter space, a single nested sampling evaluation with sufficient live points to reliably estimate the Bayesian evidence requires approximately 100 to 500 CPU-hours. Recent advances in GPU-accelerated nested sampling do offer speedups for individual evidence evaluations \cite{prathaban2025gpu, yallup2025highdim}, and differentiable GPU frameworks are increasingly enabling efficient Bayesian inference across cosmology \cite{leeney2025jaxbandflux, leeney2025salt3}. However, these methods still require rerunning for each design configuration. In contrast, cNBRE provides amortised evaluation across the entire design space. Our analysis suggests a break-even point of approximately 50--100 design evaluations, after which cNBRE becomes substantially more efficient than even GPU-accelerated traditional approaches.

We note that while Gaussian Process (GP) surrogates are often used for efficient design optimisation (e.g., in Bayesian Optimization), they typically struggle when the observation space (the simulator output) is high-dimensional, such as time-ordered data or high-resolution spectra \cite{cranmer2020frontier, miller2021truncated}. cNBRE leverages the feature-extraction capabilities of deep neural networks to learn directly from high-dimensional raw data, avoiding the need for hand-crafted summary statistics that might discard information relevant to the design trade-offs.

\subsection{Limitations and Validation}
Despite its advantages, our approach has limitations that must be carefully considered. A fundamental constraint is the reliance on the accuracy of the forward simulator. As with any simulation-based method, if the simulator fails to capture critical systematic effects or realistic noise properties, the network will yield biased design inferences \cite{schmitt2024detecting}. The optimisation is only as valid as the physics encoded in the training data.

Validating an amortised estimator across a continuous design space is a significant challenge. In this work, rather than relying solely on expensive ground-truth evidence calculations, we validated the estimator by verifying its consistency with physical expectations. In Experiment 1, the network successfully recovered the expected sigmoid relationship between observation time $\tau$ and detection probability, recovering the expected sigmoid relationship with the correct limiting behaviours ($P_{\mathrm{det}} \to 0$ at low $\tau$ and $P_{\mathrm{det}} \to 1$ at high $\tau$). Similarly, in Experiment 2, the network independently recovered the four-fold rotational symmetry expected from the physics of a dipole antenna, predicting a $\sim$20 percentage point variation in detection probability. These results strongly suggest that the network has learned the underlying physical structure of the likelihood rather than memorising spurious patterns, a risk that is further mitigated by the use of an infinite stream of training data from the simulator.

We argue that for design optimisation, the \textit{relative} ranking of designs is more critical than the precise calibration of the absolute Bayes factor. The qualitative features of the detection landscape, such as the shape of the detection probability curve, are robust indicators for decision-making. Nevertheless, a robust deployment should include spot-checks against gold-standard nested sampling runs at a few critical design points (e.g., the predicted optimum) to verify the absolute scale of the evidence.

\subsection{Broader Applicability}
The principles of cNBRE are broadly applicable beyond radio cosmology. Neural network approaches have proven valuable across 21-cm cosmology, from signal emulation \cite{dorigojones2025kan} and simulation-based inference \cite{saxena2024sbi} to the design optimisation framework presented here. Any field that relies on complex, non-linear forward models and faces expensive experimental trade-offs could benefit from this approach. Simulation-based inference has already demonstrated transformative potential in particle physics \cite{brehmer2020sbi}, neuroscience \cite{lueckmann2017flexible}, and gravitational wave astronomy \cite{bhardwaj2023peregrine}; cNBRE extends this paradigm from parameter inference to experimental design optimisation.

In the context of the REACH experiment, our results suggest that optimising geometric factors like antenna orientation could yield a $\sim$20 percentage point increase in detection probability. In a mission where the signal is buried deep in noise, such an improvement could be the deciding factor between a non-detection and a groundbreaking discovery. Furthermore, such optimisations can represent value equivalent to millions of pounds. As noted in Section~\ref{sec:intro}, facilities like the SKA (\texteuro2 billion) and LHC (\$9 billion) routinely invest heavily in upgraded hardware components for marginal sensitivity gains. Simulation-based design optimisation, which can guide optimal component selection and design decisions for a fraction of the cost, offers a high-leverage path to improved scientific return.

\subsection{Future Directions}\label{sec:future}
Future work will focus on refining the validation protocol, potentially integrating active learning to query true likelihood evaluations at regions of high uncertainty. We also plan to extend the REACH case study to a realistic multi-snapshot observation strategy, marginalising over a full year of sky rotation to assess how the orientation dependence identified here evolves with observing season. The design space would then expand to include the scheduling of observations over time and additional design parameters such as antenna polarisation. By leveraging the differentiability of the cNBRE estimator, we aim to move towards fully automated, gradient-driven instrument design, allowing the physics of the signal to directly dictate the optimal hardware configuration.

\section{Conclusion}\label{sec:conclusion}

In this paper, we presented Conditional Neural Bayes Ratio Estimation (cNBRE), a framework that extends simulation-based inference to the domain of experimental design. By training a classifier to estimate the Bayes factor via the exponential loss, cNBRE learns a direct mapping between design parameters and expected utility. While this approach incurs a substantial upfront computational cost for training, it is offset by the resulting amortised inference capability, rendering the evaluation of experimental utility continuous, differentiable, and orders of magnitude faster than traditional point-wise Monte Carlo methods.

We validated this framework using the REACH 21-cm cosmology experiment, a mission characterised by extreme spectral foregrounds. The analysis successfully recovered the expected sigmoid detection probability profile and identified complex geometric dependencies, specifically revealing an approximately 20 percentage point sensitivity variation based on antenna orientation for a single night of observation. These results demonstrate that cNBRE can resolve fine-grained features of the detection landscape that represent a computational bottleneck for standard grid-based search techniques.

Conceptually, cNBRE bears an analogy to hierarchical Bayesian modelling: just as hierarchical models learn population-level structure across related inference problems, cNBRE learns how the evidence varies across a family of experimental configurations, amortising this structure into a single network rather than solving each configuration independently.

Crucially, by directly optimising the decision-theoretic utility of detection probability rather than information gain, cNBRE provides a design metric that is strictly aligned with the primary objective of discovery experiments. Looking forward, the differentiable nature of the cNBRE utility surface provides the gradients necessary for future direct optimisation via gradient ascent, allowing for efficient exploration of high-dimensional design spaces that are currently intractable. As experiments target increasingly subtle phenomena, such as the even fainter dark ages 21-cm signal \cite{bevins2025darkages, artuc2025cosmocube}, cNBRE provides a rigorous methodology to maximise the probability of scientific discovery in low signal-to-noise regimes, ensuring hardware configurations are optimised for the decision that matters most: detection.

\section*{Acknowledgment}

SAKL is supported by the UKRI ERC Guarantee scheme.

\section*{Data Availability}

The cNBRE codebase is publicly available at \url{https://github.com/samleeney/cNBRE_}. The REACH-specific simulation code may be made available upon reasonable request to the REACH collaboration.

\bibliographystyle{IEEEtran}


\end{document}